\documentclass[namedreferences]{SolarPhysics}
\usepackage[optionalrh,solaenum]{spr-sola-addons} 
\usepackage{epsfig}                     
\usepackage{graphicx}                    
\usepackage{color}                       
\usepackage{url}                         
\usepackage{color}
\usepackage{bigmath}

\renewcommand{\Vec}[1]{{\bfmath{#1}}}

\begin{document}

\begin{article}

\begin{opening}

\title{The large longitudinal spread of solar energetic particles during the January 17, 2010 solar event}

%
\author{N.~\surname{Dresing}$^{1}$\sep
      R.~\surname{G\'omez-Herrero}$^{1}$\sep
      A.~\surname{Klassen}$^{1}$\sep
      B.~\surname{Heber}$^{1}$\sep
	  Y.~\surname{Kartavykh}$^{2,3}$\sep
	  W.~\surname{Dr\"oge}$^{2}$
       }

%

%
  \institute{$^{1}$ Institut f\"ur Experimentelle und Angewandte Physik, Christian-Albrechts-Universit\"at zu Kiel, Germany\\
  email: \url{dresing@physik.uni-kiel.de}\\
             $^{2}$ Institut f\"ur Theoretische Physik und Astrophysik, Universit\"at W\"urzburg, Germany\\
             $^{3}$ Ioffe Physical-Technical Institute, St-Petersburg, Russia
             }

\begin{abstract}
We investigate multi-spacecraft observations of the January 17, 2010 solar energetic particle event.
Energetic electrons and protons have been observed over a remarkable large longitudinal range at the two STEREO spacecraft and SOHO suggesting a longitudinal spread of nearly 360 degrees at 1\,AU.
The flaring active region, which was on the backside of the Sun as seen from Earth, was separated by more than 100 degrees in longitude from the magnetic footpoints of each of the three spacecraft.
The event is characterized by strongly delayed energetic particle onsets with respect to the flare and only small or no anisotropies in the intensity measurements at all three locations.
The presence of a coronal shock is evidenced by the observation of a type II radio burst from the Earth and STEREO~B.
In order to describe the observations in terms of particle transport in the interplanetary medium, including perpendicular diffusion, a 1D model describing the propagation along a magnetic field line (model 1) \cite{Droege2003} and the 3D propagation model (model 2) by \cite{Droege2010} including perpendicular diffusion in the interplanetary medium have been applied, respectively.
While both models are capable of reproducing the observations, model 1 requires injection functions at the Sun of several hours.
Model 2, which includes lateral transport in the solar wind, reveals high values for the ratio of perpendicular to parallel diffusion.
Because we do not find evidence for unusual long injection functions at the Sun we favor a scenario with strong perpendicular transport in the interplanetary medium as explanation for the observations.
\end{abstract}

%

\end{opening}

%
\section{Introduction}\label{s:intro} 
Energetic particle observations in the interplanetary medium provide fundamental information on acceleration processes and transport mechanisms.
The well known picture of impulsive and gradual events developed by \inlinecite{Reames1999} orders solar energetic particle (SEP) events into two different types.
While impulsive SEP events were attributed to acceleration in a small-sized reconnection site yielding to narrow longitudinal distributions of SEPs at 1\,AU around the nominal magnetic connection, gradual events were attributed to acceleration in extended sources as coronal or interplanetary shocks. 
These were supposed to be capable of generating angular SEP distributions of more than 100 degrees from the flare site 
\cite{Cane2003,Cliver2005,Kallenrode1993a}.
Using single spacecraft observations, \inlinecite{Cliver1995} reported an 180 degrees wide longitudinal spread of SEPs associated with a backside flare and attributed the observations to a coronal shock, which may have extended up to 300 degrees. 
Interplanetary shocks at 1\,AU have been suggested to be as large as 180 degrees in longitude \cite{Torsti1999,Cliver1996}, providing large acceleration regions as well.
Multi-spacecraft investigations have been used for an improved characterization of SEP events. 
Simultaneous observations by Helios, or Ulysses and the Earth caused a debate on the presence of efficient particle transport perpendicular to the magnetic field. 
A number of authors \cite{Wibberenz2006,Dalla2003,Dalla2003a,Cane2003,McKibben2001} presented observational and modeling evidence supporting perpendicular transport in the interplanetary medium.
On the other hand \inlinecite{Sanderson2003} showed particle anisotropies observed at the onset of large SEP events, which were field-aligned with small or zero flow transverse to the magnetic field, being too small to account for perpendicular transport.
Additional observational facts like the particle drop-out events \cite{Mazur2000} and nearly scatter free propagating electron spikes \cite{Klassen2011} are difficult to explain in a global transport picture including efficient perpendicular diffusion and point to a complex structure of the interplanetary magnetic field (e.g. \inlinecite{Borovsky2008}; \inlinecite{Dunzlaff2010}).

For an accurate determination of SEP properties as the angular spread, multi-point observations are invaluable. 
The STEREO mission launched in 2006 has been optimized to perform such multi-point observations.
Equipped with remote sensing and in-situ instruments, it allows to identify a possible source region of SEPs for nearly all events observed.
This was not possible prior to this mission due to the lack of remote sensing instrumentation aboard the HELIOS and Ulysses spacecraft.
Another major advantage is that the radial distance for the STEREO spacecraft and the Earth is approximately the same, which minimizes the influence of a varying radial distance on such an analysis.

The twin spacecraft of the STEREO mission, equipped with identical instrumentation, perform heliocentric orbits following the motion of the Earth in the ecliptic plane. 
With one spacecraft moving ahead of the Earth (STEREO~A) and the other trailing behind (STEREO~B), the longitudinal separation between both spacecraft grows at 45 degrees per year while the radial distance to the Sun stays nearly the same at $\sim$1\,AU.
The well separated STEREO spacecraft offer an excellent platform for multi-point studies of wide-spread SEP events.
\begin{figure}[h!] 
\centerline{\includegraphics[width=0.68\textwidth, clip=true, trim = 0mm 0mm 0mm 0mm]{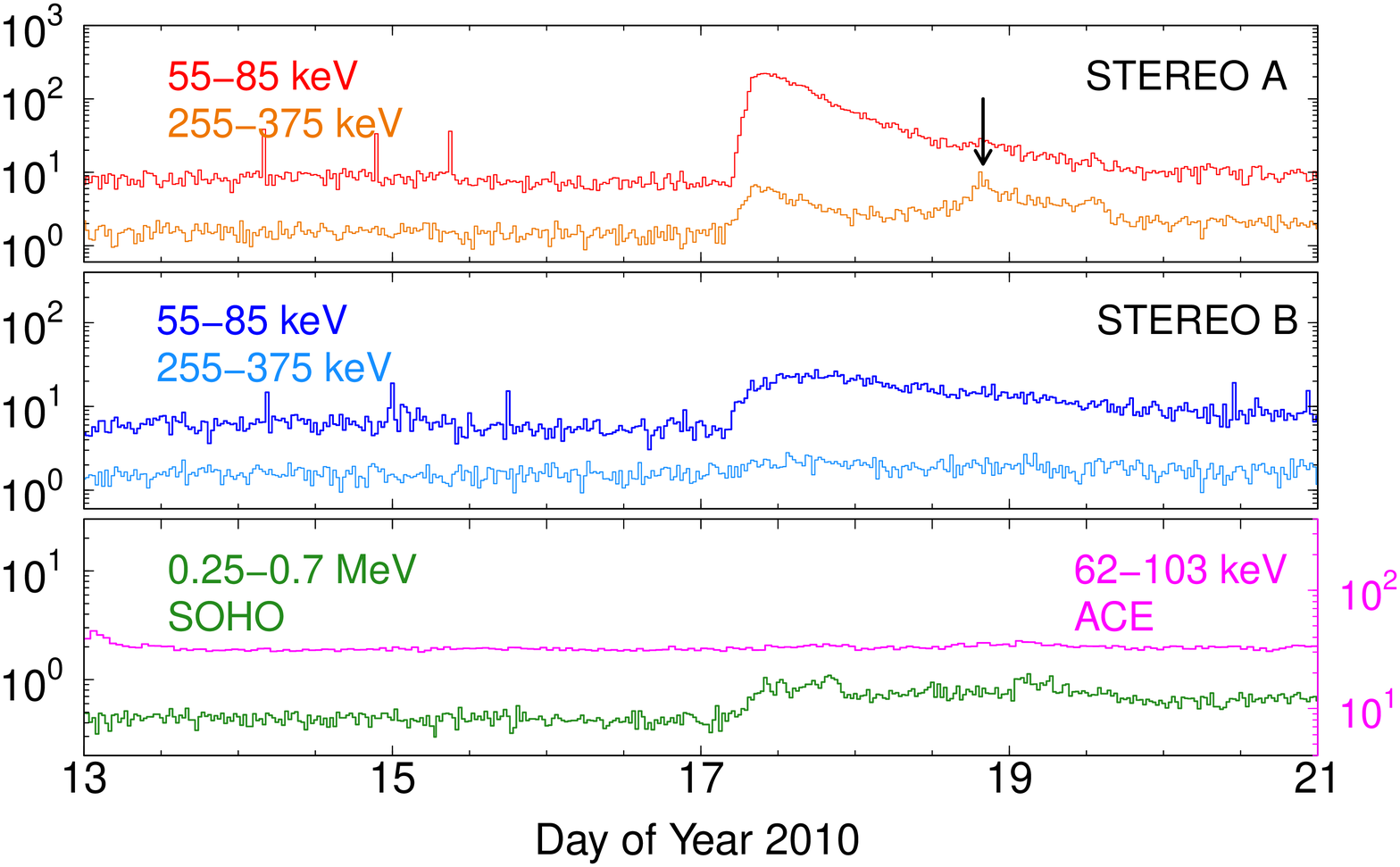}
\includegraphics[width=0.35\textwidth, clip=true, trim = -10mm -20mm 0mm 0mm]{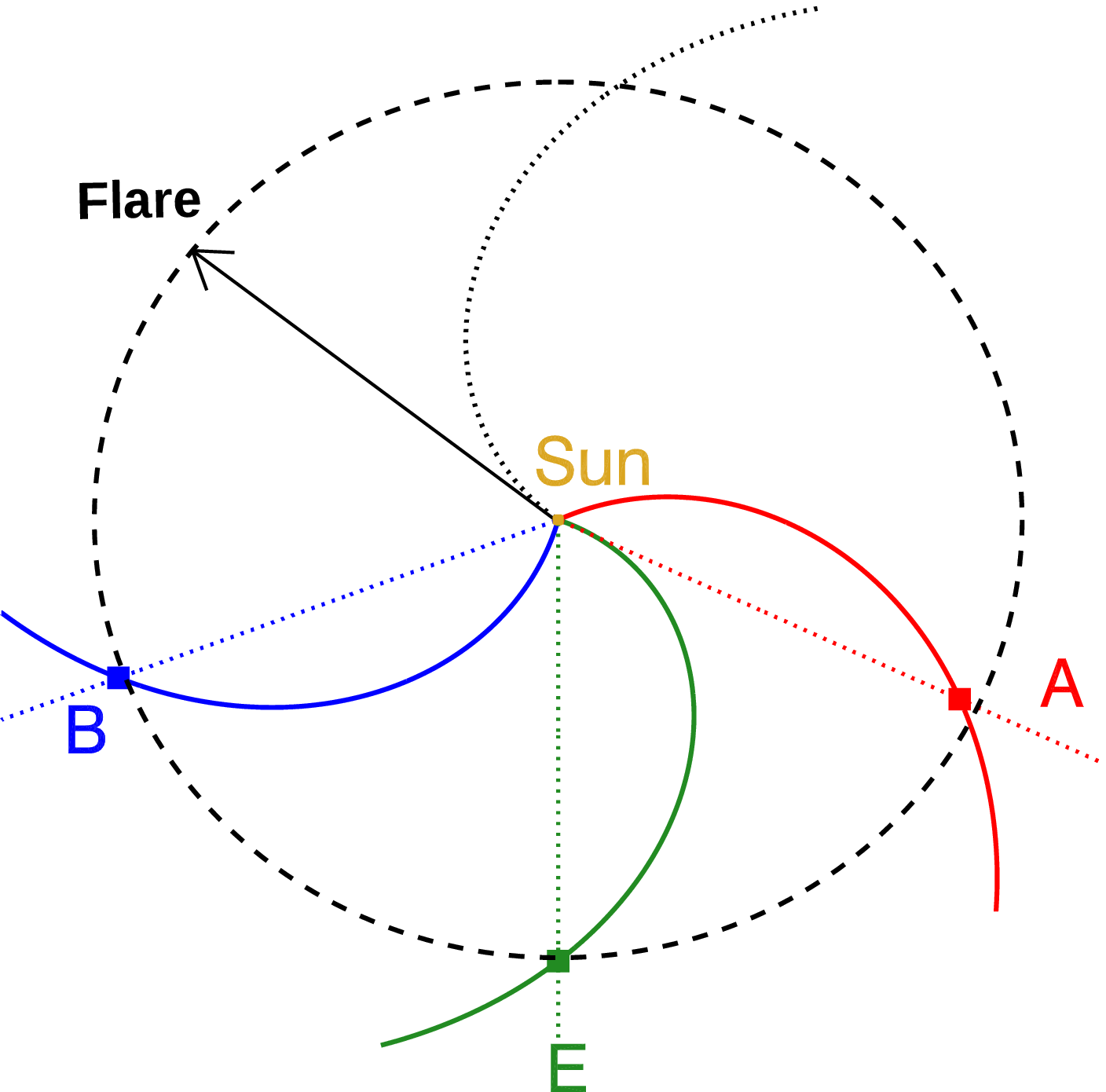}} 
 \caption{Left: Electron observations by the SEPT sunward pointing telescopes aboard STEREO~A (top), and STEREO~B (center), and SOHO/EPHIN and ACE/EPAM (bottom) showing the solar energetic particle event on January 17, 2010. The second increase observed by STEREO~A, marked by the arrow, is due to strong ion contamination.
 Right: Sketch of the longitudinal configuration of the STEREO and SOHO spacecraft with respect to the flare longitude (represented by the black arrow). The spacecraft longitudinal positions are given by colored dotted lines intersecting the dashed 1\,AU circle, where blue corresponds to STEREO~B, red to STEREO~A, and green to SOHO / ACE, respectively. 
The magnetic field lines connecting the Sun and the spacecraft are given by the colored spirals and correspond to the measured in-situ solar wind speed. The black spiral represents the nominal magnetic field line originating from flare position.} \label{fig:intro}
 \end{figure}%
On January 17, 2010 the spacecraft were separated by 134 degrees, with the Earth in between, when instruments at all three locations measured enhanced electron and proton fluxes.
We identified the source region on the backside of the Sun as seen from Earth unambiguously at Carrington longitude 54 and -25 degrees latitude.
Thus the longitudinal spread of the energetic particles at 1\,AU during this event is expected to be larger than 300 degrees. 
Figure \ref{fig:intro} (left) shows measurements at STEREO and ACE / SOHO in approximately the same energy range from $\sim$50-100\,keV and $\sim$250-400\,keV.
The second increase observed by STEREO~A, marked by an arrow, is not real but due to ion contamination.
The bottom panel also shows ACE/EPAM (LEFS 60) electron measurements in the range of 62-103\,keV. 
Here the event is masked by the high background but appears with more than two sigma when background subtraction is applied (not shown), confirming the SOHO/EPHIN observations.
Note, that for the DE channels the event is not seen even with background subtraction. 
On the right hand side a sketch shows the longitudinal configuration of the spacecraft with respect to the flare position.
In what follows, we discuss the instrumentation as well as the observations, and compare the observations with results of a solar energetic particle transport code \cite{Droege2010}.
%
%
%
\section{Instrumentation}\label{s:instru}\label{instru}
In this work, we use energetic particle measurements by the Low Energy Telescope (LET, \inlinecite{Mewaldt2007}), the High Energy Telescope (HET, \inlinecite{VonRosenvinge2007}), and the Solar Electron and Proton Telescope (SEPT, \inlinecite{Muller-Mellin2007}) contained in the IMPACT instrument suite \cite{luhmann2007} aboard the STEREO spacecraft.
The SEPT instrument measures electrons in the range of 30-400\,keV and nuclei from 60-7000\,keV/n, each in four looking directions, which are to the north (NORTH), to the south (SOUTH), along the Parker spiral to the Sun (SUN), and away from the Sun (ANTISUN), respectively. 
The SECCHI investigation \cite{Howard2008} provides remote sensing observations of the Sun in extreme ultra violet (EUVI, \inlinecite{Wuelser2004}) as well as coronagraphic observations (COR1 and COR2 instruments), which allow to link in-situ observations with the associated regions of solar activity. 
In this work, the STEREO observations are complemented by SOHO and ACE measurements. 
The Electron Proton and Alpha Monitor (EPAM) on board ACE measures the flux and direction of ions greater than 0.2\,MeV to 93\,MeV \cite{Gold1998}.
The Electron Proton Helium Instrument (EPHIN, \inlinecite{Muller-Mellin1995}) on board SOHO measures energetic electrons and protons, remote sensing observations are performed by the Extreme-Ultraviolet Imaging Telescope (EIT, \inlinecite{Delaboudiniere1995}) and the Large Angle and Spectrometric Coronagraph (LASCO, \inlinecite{Brueckner1995}). 
Signatures of radio bursts are detected with the WAVES instrument aboard WIND \cite{Bougeret1995} and the STEREO/WAVES instruments \cite{Bougeret2008}.
Solar wind plasma and interplanetary magnetic field data are provided by the STEREO/PLASTIC \cite{Galvin2008} and STEREO/MAG  \cite{Acuna2007} instruments, respectively.
%
%
\section{Observations}\label{s:obs} 
In this section, we first present the remote sensing observations (Section \ref{remote}), followed by the in-situ measurements (Section \ref{insitu}), and finally we describe the interplanetary (plasma and magnetic field) context in Section \ref{IPcontext}.
%
\subsection{Remote Sensing Observations}\label{remote}
\begin{figure}[h!] 
\centerline{\includegraphics[width=\textwidth]{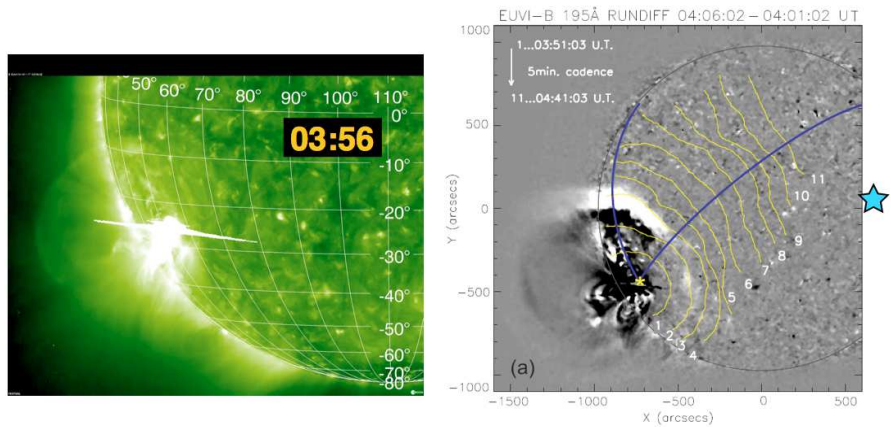}}
 \caption{STEREO B EUVI 195\,{\AA} observations. Left: The flare maximum on January 17, 2010 at 3:56\,UT. Right: 5-min running difference image showing the EUV wave (adapted from \protect\inlinecite{Veronig2010}, reproduced by permission of the AAS). The yellow lines mark the visually identified outer edges of the wavefronts. The blue star slightly outside the figure represents the position of the STEREO~B magnetic footpoint.} \label{fig:flare}
 \end{figure}

\begin{figure}[h!] 
\centerline{\includegraphics[width=\textwidth]{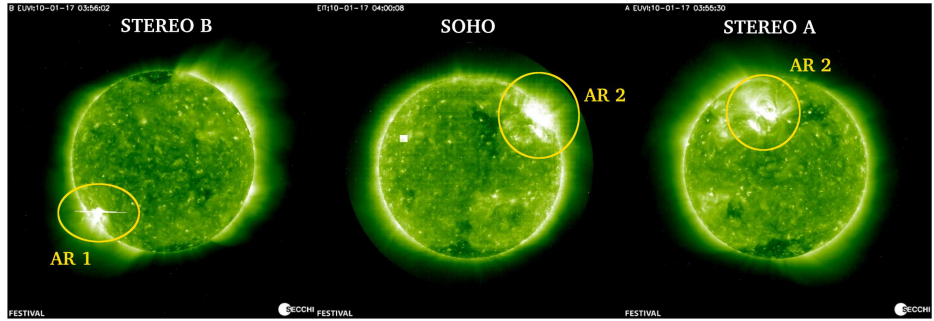}}
 \caption{From left to right: EUV observations by STEREO~B, SOHO and STEREO~A during the flare time on January 17. Two different active regions were observed at the Sun, which are labeled with 'AR 1' (NOAA AR 11039) and 'AR 2' (NOAA AR 11040), respectively. A flare was only seen in AR 1 by STEREO~B.} \label{fig:flare_all}
 \end{figure}

On January 17, 2010 active region (AR) 11039, which recurred in the following rotation as AR 11041, produced a flare.
The flare was first detected with the 171\,{\AA} band of the EUVI instrument aboard STEREO~B at 3:49\,UT.
As shown in Figure \ref{fig:flare} (left), the eruption site was located at E59 S25 as seen from STEREO~B, which corresponds to $\sim$37 degrees behind the east limb as seen from Earth. 
Only one other AR has been observed at this time from STEREO~A and the Earth (see AR2 in  Figure \ref{fig:flare_all}) showing no significant activity during this period.
Extreme ultraviolet (EUV) and coronograph observations by STEREO~B show a large-scale dome-shaped expanding coronal EUV wave with perfectly connected off-limb and on-disk signatures as reported by \inlinecite{Veronig2010} and \inlinecite{Grechnev2011}. 
An EUV wave propagating towards the north-western direction is shown in the difference image in Figure \ref{fig:flare} (right) adapted from \inlinecite{Veronig2010}. 
The yellow lines show the propagation of the front in five minute steps. The blue star slightly outside the figure represents the position of the STEREO~B magnetic footpoint.
%
 \begin{figure} [h!] 
\centerline{\includegraphics[width=\textwidth]{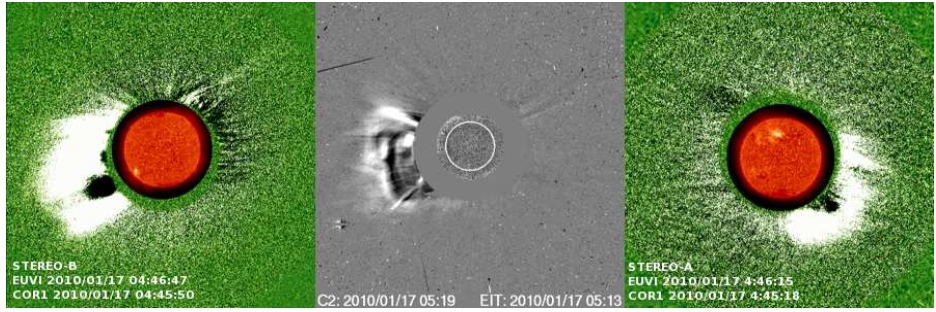}}
 \caption{STEREO and SOHO observations of a CME on January 17, 2010 provided by the CME catalogs http://cor1.gsfc.nasa.gov/catalog/ and http://cdaw.gsfc.nasa.gov/CME\_list/. Left: STEREO~B EUVI and COR1, center: SOHO/LASCO, right: STEREO~A EUVI and COR1. To increase the visibility of the CME, the coronograph observations are shown as difference images.}\label{fig:cme}
 \end{figure}%
While the flare and EUV wave were observed only by STEREO~B, a CME, directed slightly southwards, was observed by all three viewpoints as displayed in Figure \ref{fig:cme} (adapted from the COR1 CME catalog\footnote{http://cor1.gsfc.nasa.gov/catalog/} and the SOHO/LASCO catalog\footnote{http://cdaw.gsfc.nasa.gov/CME\_list/}). As reported in the STEREO COR1 CME catalog, the CME was seen in SE direction from STEREO~B and in SSW direction from STEREO~A, respectively, with first appearance in the COR1 fields of view at 4:10\,UT.
The SOHO/LASCO CME catalog reports a partial halo CME appearing in the C2 field of view at 4:50\,UT with a linear fit speed of 350\,km s$^{-1}$, a central angle of 114 degrees, and an angular width of 126 degrees.

\begin{figure} 
\centerline{\includegraphics[width=0.8\textwidth, clip=true, trim  = 0mm 35mm 0mm 0mm]{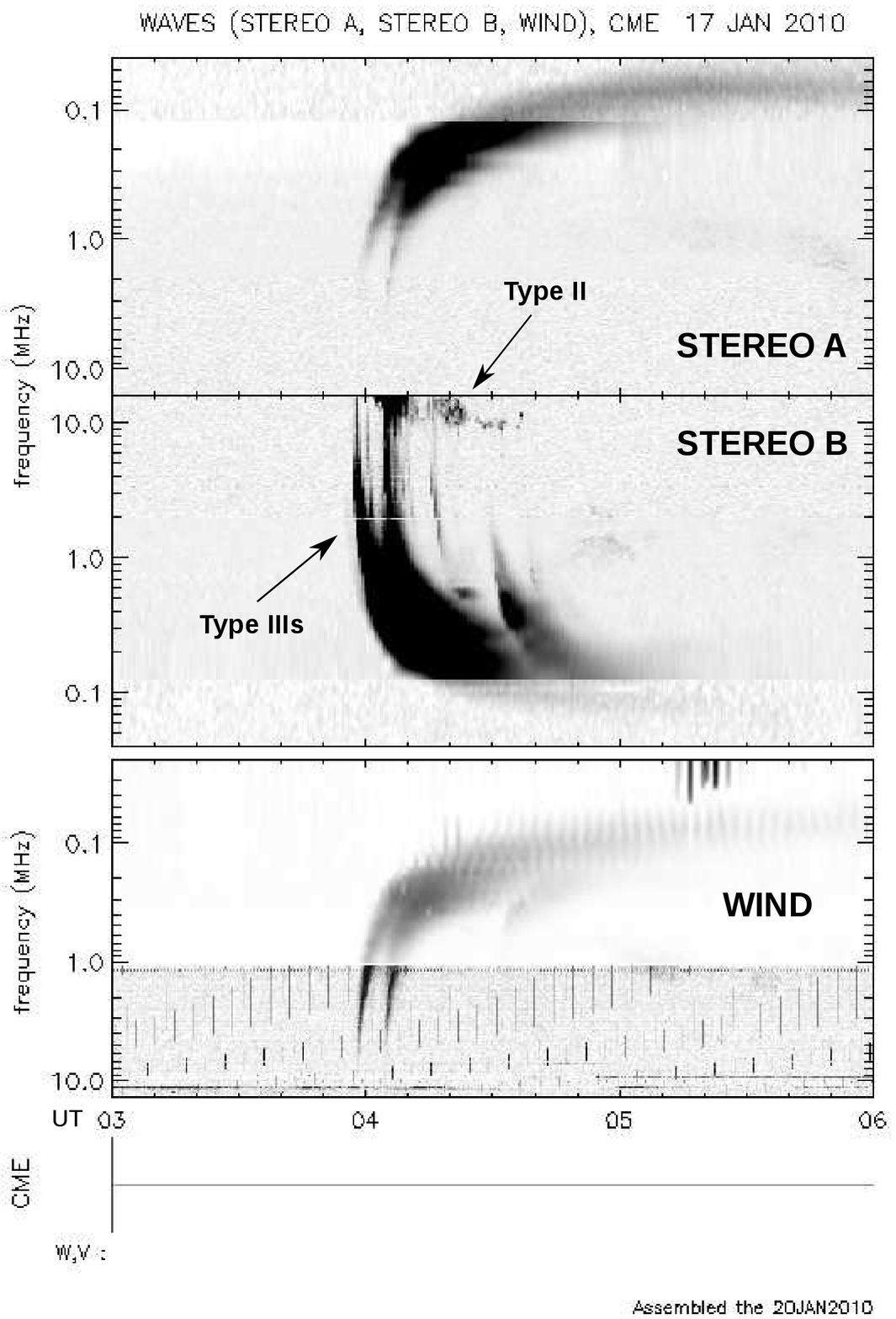}}
 \caption{Radio spectrograms recorded by the SWAVES instruments aboard STEREO A (top), STEREO~B (middle) and WIND/WAVES (bottom) provided by the Meudon Radio Monitoring Survey (http://secchirh.obspm.fr/select.php).}\label{fig:radio}
\end{figure}

Figure \ref{fig:radio} shows dynamic spectra of radio waves detected aboard STEREO~A (top), STEREO~B (center), and WIND (bottom), provided by the Meudon Radio Monitoring Survey\footnote{http://secchirh.obspm.fr/select.php}.
Type III radio bursts were observed between 3:58-4:28\,UT by all three spacecraft accordingly with the flare time at 3:49\,UT.
Its high frequency part in the STEREO~A and WIND spectrograms shows a clear attenuation.
This is also in agreement with the relative positions of the AR 11039 and presumably of the type III sources) of about 77 degrees behind the west limb for STEREO A, and 37 degrees behind the east limb for WIND.
A type II radio burst indicating the presence of a coronal/IP shock was detected at STEREO~B from 4:02 to 4:37\,UT, visible in Figure \ref{fig:radio}, second panel, and has also been reported by the ground based radio observatory HiRAS from 3:51 to 3:58\,UT in the range of 310-80\,MHz. \inlinecite{Grechnev2011} concluded that "the EUV wave most likely was a near-surface trail of a large-scale coronal MHD wave", and that "the shock was weak to moderate".
The Figure shows no further radio activity detected hours before and after the event. 
Consequently, the combination of radio and EUV observations by the three viewpoints discards the presence of another source region apart from the one (AR11039) clearly seen on the visible disk for STEREO~B.
Other eruptive events from the same AR can be excluded by the absence of further type III bursts for more than seven hours around the investigated event.
\begin{figure} 
\centerline{\includegraphics[width=1.05\textwidth]{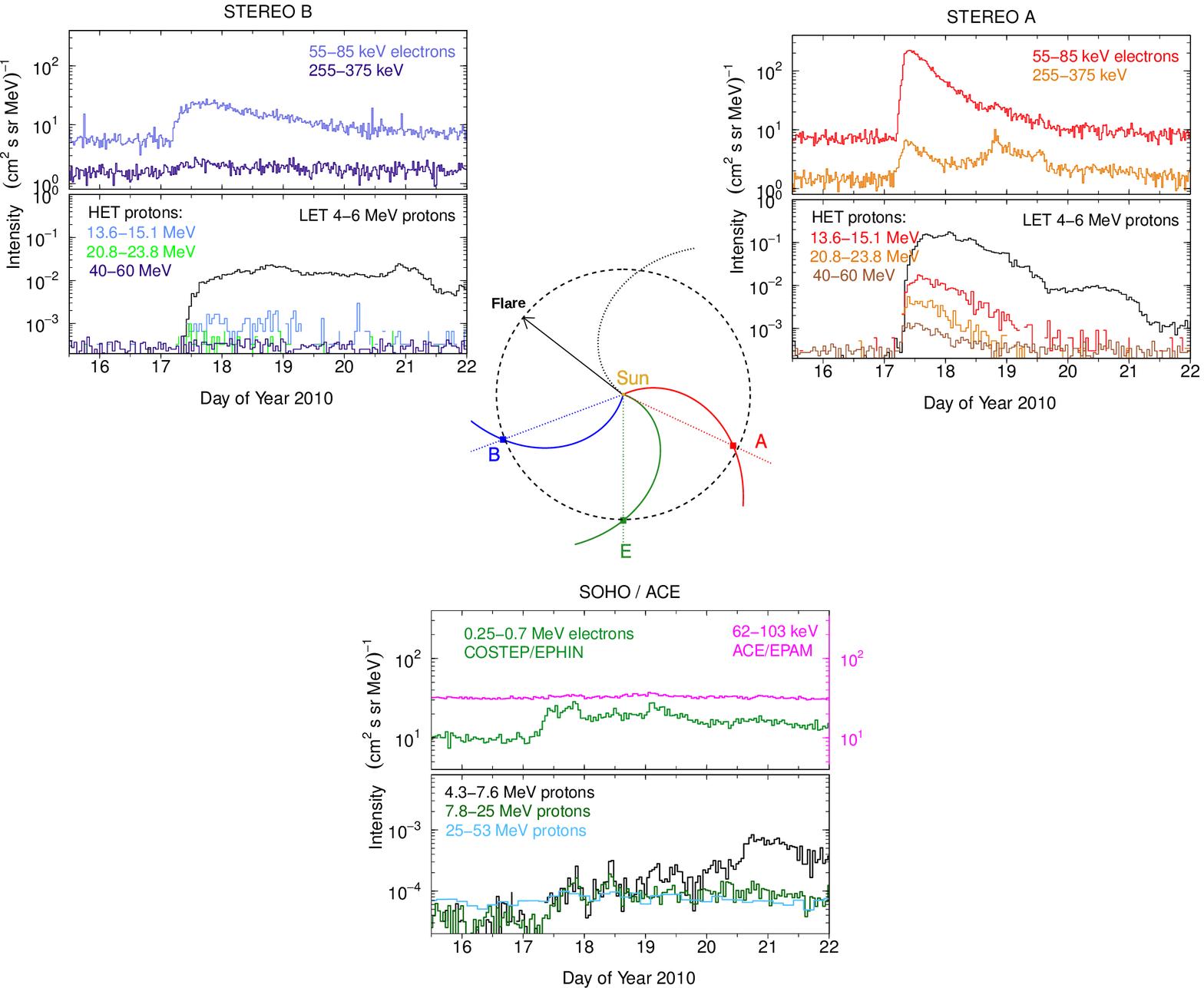}}
 \caption{Sketch of the longitudinal configuration of the STEREO and SOHO spacecraft with respect to the flare longitude as in Figure \ref{fig:intro}. The sketch is surrounded by electron (top panels) and proton (bottom panels) measurements by the SEPT, LET and HET instruments aboard STEREO~B (left), STEREO~A (right), and SOHO/EPHIN (bottom).}\label{fig:config}
\end{figure}%
%
\subsection{In-situ Observations}\label{insitu}
On January 17, 2010 STEREO~B was located 69.2 degrees behind the Earth and SOHO longitude, and STEREO~A was 64.7 degrees ahead of it.
The configuration of the three spacecraft and the flare position is sketched in Figure \ref{fig:config} (center). 
Dashed colored lines and the black arrow mark the longitudinal positions of the spacecraft and the flare, respectively. 
The spiral lines represent the connecting magnetic field lines between the spacecraft and the Sun using the measured solar wind speed, and the black spiral is the nominal magnetic field line connecting to the AR. 
The dashed black circle indicates a 1\,AU distance to the Sun.
The longitudinal separation between the spacecraft magnetic footpoint, consistent with the solar wind speed measured in-situ, and the flare position is about 113 degrees for STEREO~B, 117 degrees for STEREO~A, and 161 degrees for SOHO. The top panels of the three plots surrounding the sketch in Figure \ref{fig:config} show electron time profiles by the sunward pointing SEPT telescopes aboard STEREO~B (left) and STEREO~A (right) in the energy range from 65 to 85\,keV, and by the SOHO/EPHIN instrument (bottom) in the lowest available energy channel ranging from 250 to 700\,keV. Energetic proton measurements in the range from 4 to 60\,MeV taken by the LET and HET instruments aboard the two STEREOs and protons in the range from 4 to 25\,MeV measured by the SOHO/EPHIN instrument are displayed in the bottom panels.

The particles arrive at all three spacecraft delayed with respect to the flare onset (which is taken as first optical (EUV) flare observation at 3:41\,UT) with at least 49 minutes (STEREO~B) delay for $<$100\,keV electrons.
\begin{figure} 
\centerline{\includegraphics[width=\textwidth, clip=true, trim  = 0mm 0mm 0mm 0mm]{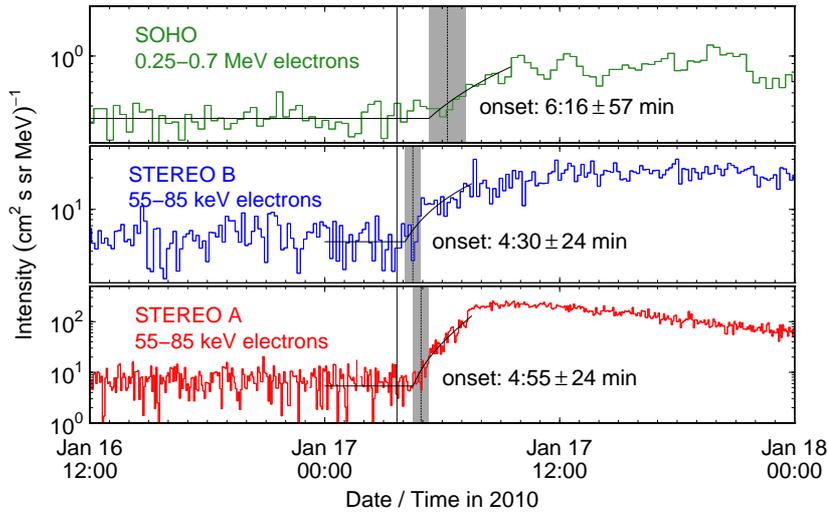}}
\caption{Electron measurements and onset times (vertical dashed lines) at STEREO~A (bottom), STEREO~B (center), and SOHO (top). The gray shadow represents the range of uncertainty and the solid black line marks the flare onset time (shifted by 8 minutes light travel time). Due to poor statistics, the data had to be averaged before determining the onset times by 4 minutes for STEREO~A, 10 minutes for STEREO~B, and 20 minutes for SOHO.}\label{fig:onsets}
\end{figure}%
Figure \ref{fig:onsets} shows electron measurements at the three spacecraft. The electron onset times of the event are indicated by dashed lines in the center of the gray shadow. 
The onset time of the flare at 3:41\,UT is presented by the solid black line and has been corrected by eight minutes light travel time. 
The gray shadow illustrates the range of uncertainty for the specific onset time defined as the range between a lower and an upper limit for the onset time.
The upper limit is the first time when the electron flux exceeds the pre-event mean flux by at least three standard deviations and continues above that value for at least two consecutive intervals. 
The lower limit was determined by fitting the data with the function
\begin{equation}
I(t)=I_0+ A \cdot e^{B(t-t_0)}
\end{equation}
with free parameters $I_0, A, B$ and the onset time $t_0$.
While the fit retained relatively stable for the STEREO~A data, it was very sensitive to the fitting window for STEREO~B and SOHO data due to the poor statistics.
Therefore, an additional inspection by eye was needed to verify the specific fit parameters. 
The fits are also included in the figure.
While the electron onset at both STEREO spacecraft is nearly simultaneously within the range of uncertainty, there is a larger delay for the SOHO electrons.
An onset analysis has also been applied to the 4-6\,MeV protons as measured by the LET instruments aboard STEREO.
In contrast to the nearly simultaneous electron onset, the protons arrive at least one hour later at STEREO~B compared to STEREO~A (not shown).
%
 %
 \begin{figure} 
 \centerline{
 \includegraphics[width=\textwidth]{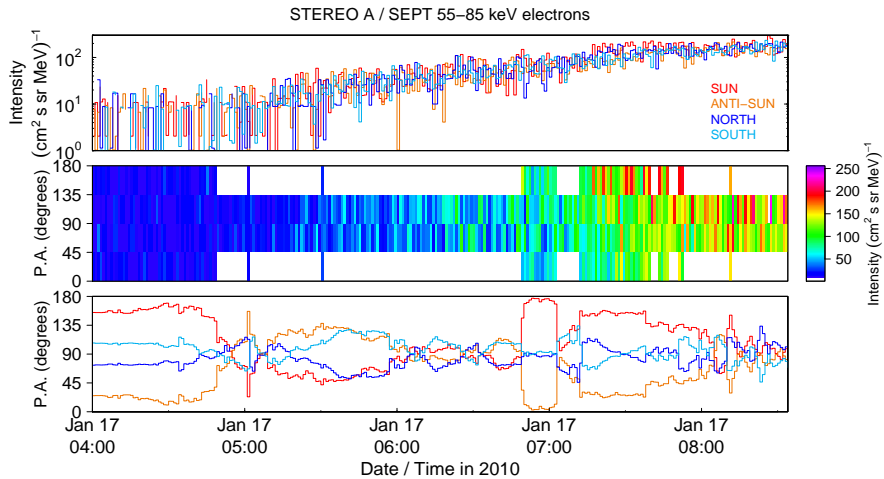}
 }
 \caption{From top to bottom: STEREO A 55-85\,keV electrons in all four looking directions of the SEPT instrument, 3D pitch angle distribution, and pitch angles of each of the four SEPT sensors. Note, that white spaces in the center panel denote sectors with no pitch angle coverage.}\label{fig:ani}
 \end{figure}%
 %
Figure \ref{fig:ani} (top) shows 65-75\,keV electrons measured by SEPT~A in all four looking directions.
The center and bottom panels of Figure \ref{fig:ani} show the 3D pitch-angle distribution and pitch angles of each SEPT sensor, respectively. 
It is evident that the pitch angle-coverage especially during the onset of the event is poor.
During these periods no clear statement on the anisotropy can be made.
Later phases of the event, when the pitch-angle coverage is better, show only a small anisotropy.
A velocity dispersion analysis for STEREO~A gave no clear results, while the STEREO~B and SOHO statistics did not allow this analysis.

Table \ref{tbl:delay} summarizes the observations made by the STEREO spacecraft and observers at Earth, SOHO, or WIND, respectively as presented so far.
Onset times of flare, radio bursts and energetic electrons (T$_{\mbox{o}}$) at the spacecraft as well as longitudinal separations between flare site and spacecraft or spacecraft magnetic footpoints ($\Delta\theta$ and $\Delta\varphi$, respectively) are listed. 
The time delays between flare and electron onset times $\Delta \mbox{t}$ as well as the travel times of the electrons along a nominal parker spiral $\mbox{t}_{\mbox{travel}}$ are included. 
Furthermore, information on SEP anisotropies and velocity dispersion are listed.
The spectral indices $\gamma$ for electrons in the range of 65 to 195\,keV for a period of 30 minutes during the maximum time of the event are listed as well.
With $\gamma$=-2.0$\pm$0.1 STEREO~B shows a harder spectrum compared to STEREO~A with $\gamma$=-2.20$\pm$0.03.
%
%
\begin{table} 
\caption{Summary of the observations. Onset times of flare, radio bursts, and first appearance of CMEs have been shifted by -8.3 minutes to correct from light travel time. If a column is empty, the feature has not been observed by the spacecraft. 
$\Delta\theta$ and $\Delta\varphi$ represent the longitudinal separations between flare and spacecraft longitude as well as between flare and spacecraft magnetic footpoint (taking into account the measured solar wind speed), respectively. A minus indicates that the flare is located eastwards of the spacecraft.
The heliographic (HG) latitude of the spacecraft is displayed as well. Electron onset times in the energy range from 55-85\,keV for STEREO and 0.25-0.7\,MeV for SOHO are indicated by $\mbox{T}_{\mbox{o}}$, and $\Delta \mbox{t}$ is the time delay between these onset times and the flare onset. $\mbox{t}_{\mbox{travel}}$ is the expected electron travel time along a nominal parker spiral of 1.2\,AU length. Anisotropy and velocity dispersion informations regard electron and proton measurements as well, while the spectral index $\gamma$ during the maximum of the event (for electrons only) is listed.}\label{tbl:delay}
\begin{tabular}{lccc}     
\hline
 							 &  STEREO~B          &  STEREO~A          &  Earth(E)/SOHO(S)      \\
 							 &                    &                    &        /WIND(W)        \\
Flare onset					 &  3:41\,UT          &  ---               &      ---               \\
Type II onset   	  		 &  $<$3:54\,UT       &  ---               &  $<$3:43\,UT (E)       \\
Type III onset         		 &  $<$3:50\,UT       &  $<$3:51\,UT       &  $<$3:50\,UT (W)       \\
CME							 &  4:02\,UT          &  4:02\,UT          &  4:42\,UT (S)          \\
$\Delta\theta$         		 &  -59$^{\circ}$     &  173 $^{\circ}$    &  -128  $^{\circ}$      \\  
$\Delta\varphi$		   		 &  -113$^{\circ}$    &  117$^{\circ}$     &  -161$^{\circ}$        \\
HG latitude                  &  3.740             &      -7.060        &   -4.753               \\
$\mbox{T}_{\mbox{o}}$ (UT)   & 4:30$\pm$24min     & 4:55$\pm$24min     &  6:16$\pm$57min (S)    \\ 
$\Delta \mbox{t}$            &     49\,min        &    74\,min         &   $\sim$2.5\,h         \\
$\mbox{t}_{\mbox{travel}}$   &  $\sim$20\,min     &    $\sim$20\,min   &   $\sim$11\,min        \\
SEP anisotropy               & unclear                 &   small       &  too poor statistics   \\
Velocity dispersion          & too poor statistics  &  unclear         &  too poor statistics   \\
Spectral index $\gamma$      &  -2.0$\pm$0.2      & -2.20$\pm$0.03     &  no overlap with       \\
                             &                    &                    &     STEREO range       \\
\hline                             
\end{tabular}   
\end{table}
%
%
\subsection{Interplanetary Context}\label{IPcontext}
\begin{figure} 
 \centerline{\includegraphics[width=0.5\textwidth, clip=true]{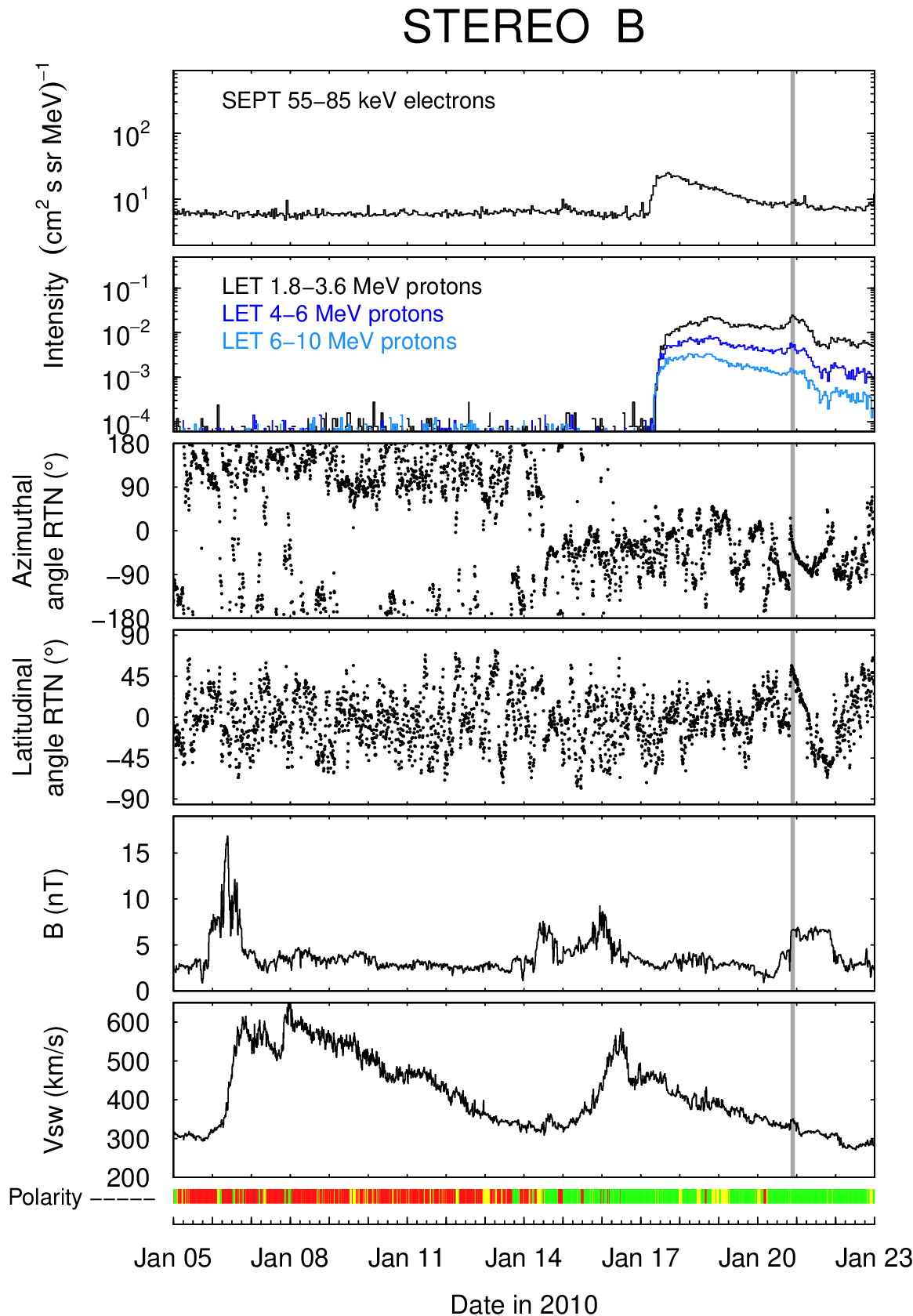}
 \includegraphics[width=0.5\textwidth, clip=true]{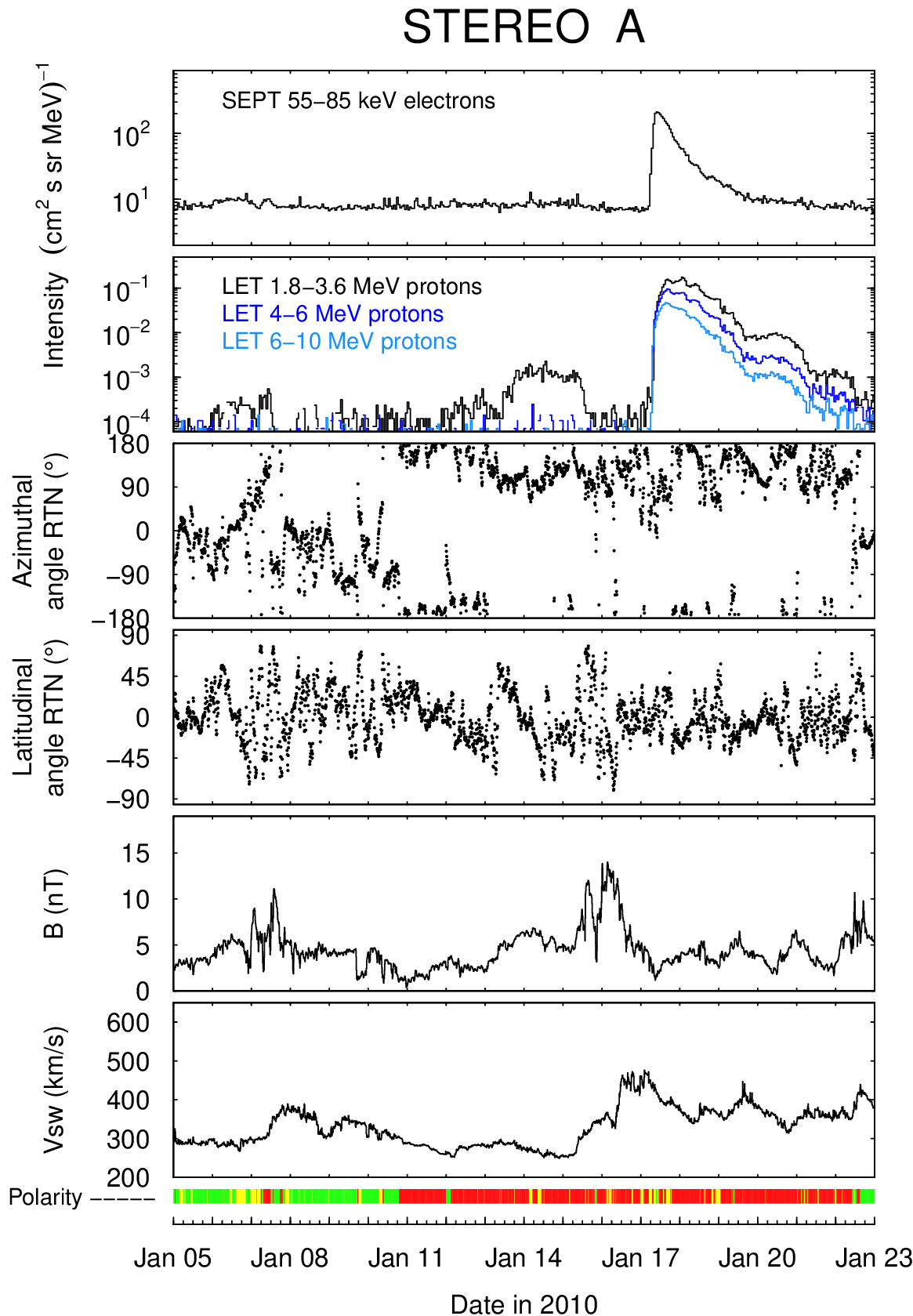}}
 \caption{In-situ measurements by STEREO~B (left) and STEREO~A (right). From top to bottom: Electron intensities, proton intensities, magnetic field azimuthal and latitudinal angles, magnetic field magnitude, and solar wind speed. The gray-shaded bar indicates a magnetic cloud measured by STEREO~B on 20 January at 20:20\,UT. The colored bar at the bottom of each plot represents the measured magnetic field polarity with negative polarity in red and positive polarity in green, respectively.}\label{fig:overview}
 \end{figure}

\begin{figure} 
 	\centerline{\includegraphics[width=0.85\textwidth, clip=true]{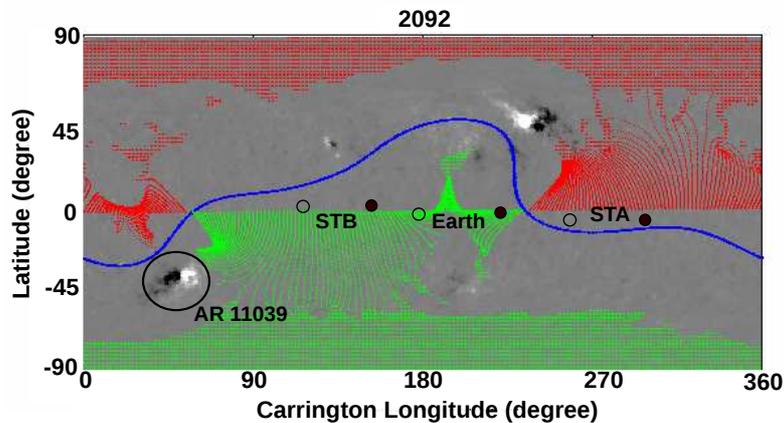}}
	 \caption{Synoptic Ecliptic-Plane Field Plot overlying a magnetogram map by the Global Oscillation Network Group (GONG) for Carrington rotation 2092. The projected positions of STEREO~B (STB), STEREO~A (STA), and the Earth have been added as open circles, while the positions of the magnetic footpoints of the spacecraft are given by filled circles always to the left.}\label{fig:gong}
\end{figure}

Figure \ref{fig:overview} shows in-situ measurements by  STEREO~B (left) and STEREO~A (right) from 1 January to 23 January 2010. 
The two top panels, presenting $<$100\,keV electrons and 4-60\,MeV protons, respectively, clearly show the 17 January 2010 SEP event. 
Below, the azimuthal and latitudinal magnetic field angles, the magnetic field magnitude, and the solar wind speed are displayed. 
The red/green colored band at the bottom of each plot indicates the in-situ magnetic field polarity where red is negative and green is positive polarity. 
During the SEP event, STEREO~B stands in a positive polarity magnetic sector, while STEREO~A and the Earth (not shown) are in a negative magnetic sector.
Figure \ref{fig:gong} shows a Potential Field Source Surface (PFSS) Carrington map overlying a magnetogram map of Carrington rotation 2092 provided by the Global Oscillation Network Group (GONG)\footnote{http://gong.nso.edu/}. 
The blue wavy line represents the neutral sheet separating the negative polarity sector, with the upper most closed magnetic field lines marked in red, from the positive polarity sector in green, respectively.
The projected positions of the spacecraft have been added as open circles while the positions of the magnetic footpoints of the spacecraft (always to the right) are represented by the filled circles.
The AR producing the 17 January event is seen in the bottom left part of the map in a positive polarity magnetic sector. 
The STEREO in-situ measurements in Figure \ref{fig:overview} show some minor compression regions caused by stream interactions but no completely evolved CIR with shocks.
The small gray bar on January 20 in the STEREO~B plot in Figure \ref{fig:overview} indicates a magnetic cloud possibly associated with the CME from the 17 January event.
According to the STEREO COR1\footnote{http://cor1.gsfc.nasa.gov/catalog/} and LASCO\footnote{http://cdaw.gsfc.nasa.gov/CME\_list/} CME catalogs, at least five CMEs were present in the interplanetary medium between the Sun and 1\,AU when the 17 January SEP event occurred.
These CMEs had mostly small angular widths and low velocities.
According to Jian et al.\footnote{http://www-ssc.igpp.ucla.edu/forms/stereo/stereo\_level\_3.html} no shocks have been observed in-situ by the STEREO spacecraft during January 2010. 

%
\section{Simulations}\label{s:sim} 
\subsection{The 3D Transport Model}
The appearance of solar particle events observed in the near-Earth environment 
reflects the combination of a number of physical processes. For particles 
originating from solar flares these processes include acceleration 
in the flaring region, some kind of lateral transport in the solar corona
to a magnetic field line connected with the observer, and transport
in the solar wind. In the undisturbed solar wind, i.e., in the absence of 
CMEs and shocks, the latter can usually be described as adiabatic motion 
along the average interplanetary magnetic field represented by an Archimedean 
spiral, and pitch angle scattering off the fluctuations superimposed on the 
field. Diffusion perpendicular to the magnetic field is caused by interactions 
with fluctuations which scatter the particles' gyrocenter from one field line 
to another and by the combined effects of parallel transport and field line 
mixing (cf. \inlinecite{Jokipii1966}, \inlinecite{Jokipii1969}, \inlinecite{Ragot1999}). Transport perpendicular to the average field 
(so-called co-rotation) also 
arises due to the action of an induced electric field, $ \Vec{E} = \Vec{V_{sw}}
\times \Vec{B}$, where $\Vec{V_{sw}}$ is the solar wind velocity and $\Vec{B}$
denotes the magnetic field. Depending on particle
species and energy, also convection and energy losses in the interplanetary
medium can be of importance for the transport process. 
In the most general case the quantitative treatment of the evolution of the 
particle's gyro-averaged phase space density $f(\Vec{r},\mu,p,t)$
(which is proportional to the observed particle flux) at the location $\Vec{r}$ 
can be cast into the form (cf., \inlinecite{Droege2010} for details):
\begin{eqnarray}
  {\partial f\over\partial t} = -~\mu v \Vec{b}\cdot{\mbox{\boldmath $\nabla$}}f
  - {1 - \mu^2\over 2 L} v {\partial f\over\partial \mu} \ 
  + {\partial\over\partial \mu} \biggl( D_{\mu\mu}(\mu) {\partial f\over\partial         \mu}\biggr)    + \ q(\Vec{r},\mu,p,t) \hspace{1cm} (2\rm{a}) 
\nonumber \\[3.0ex] 
  - \ (\Vec{V_{sw}} + \Vec{V_{d}})\cdot{\mbox{\boldmath $\nabla$}}f 
  \ - \ \biggl[{\mu(1 - \mu^2)\over 2} ({\mbox{\boldmath $\nabla$}}\cdot\Vec{V_{sw}}
  -~3 \Vec{b}\Vec{b}\:{\bf :}\:{\mbox{\boldmath $\nabla$}}\Vec{V_{sw}})
         \biggr]~{\partial f\over\partial \mu} \hspace{1cm} (2\rm{b}) \nonumber
\nonumber \\[2.0ex] 
   \ + \ \biggl[{1 - \mu^2\over 2} ({\mbox{\boldmath $\nabla$}}\cdot\Vec{V_{sw}}
   - \Vec{b}\Vec{b}\:{\bf :}\:{\mbox{\boldmath $\nabla$}}\Vec{V_{sw}})
   \ + \ \mu^2 \Vec{b}\Vec{b}\:{\bf :}\:{\mbox{\boldmath $\nabla$}}\Vec{V_{sw}}
       \biggr]~p {\partial f\over\partial p} \hspace{1cm} (2\rm{c}) 
\nonumber \\[2.5ex] 
    \ + \ \ {\mbox{\boldmath $\nabla$}}\cdot(\Vec{K_\bot}{\mbox{\boldmath $\nabla$}}f). \hspace{1cm} 
    (2\rm{d}) \nonumber
\end{eqnarray}\\[-2.0ex]        \setcounter{equation}{2}  %
Here $v$ is the particle speed, $\mu=\cos\vartheta$ the particle pitch angle cosine, $t$ the time, and $\Vec{b} = \Vec{B}/B$ a unit vector in the direction of the
average magnetic field $\Vec{B}$. The focusing of particles due to the
divergence of the magnetic field is described by $L(\Vec{r}) = -( \Vec{b}\cdot{\mbox{\boldmath $\nabla$}}~\rm{ln}~\it{B}(\Vec{r}))^{-1}$,
and the stochastic forces are 
taken into account through the pitch angle diffusion coefficient $D_{\mu\mu}(\mu)$
and the tensor $\Vec{K_\bot}$ which describes the diffusion of particles in the two dimensions perpendicular to the average magnetic field direction . The injection of particles close to the Sun is given by $q(\Vec{r},\mu,p,t)$. 

As analytical solutions of transport equations for the pitch-angle dependent
phase space density are not known, numerical methods have to be applied. Finite-differences (FD) schemes have been used for this purpose since the 1980s, more recently also Monte Carlo (MC) simulations were employed to solve the corresponding stochastic differential equations (SDE). 

In this work we use two different techniques to model the particle transport (in the following called model 1 and model 2, respectively). Both approaches assume an Archimedean spiral magnetic flux tube connecting the Sun and the spacecraft, consistent with the solar wind speed measured in-situ. Also, they assume that the effects of energy losses 
and of convection are small for the near-relativistic electrons considered here, and can be neglected. 

\subsection{Model 1 - No Perpendicular Diffusion in the Solar Wind}
The first model is based on a finite-differences scheme (e.g., \inlinecite{Droege2003}). This model also neglects the effects of diffusion perpendicular to the average magnetic field and instead assumes that there is no variation across the magnetic field, and that the respective solutions are identical in neighboring flux tubes. In this case all relevant effects are contained in Equation~2a,
which is solved for a fixed energy, corresponding to the midpoint of the energy interval under consideration. The lateral spread of the electrons is assumed
to occur in the corona, and the injection on the field line connected to
the observing spacecraft at a distance of $r$=0.05 AU is described by $q(\Vec{r},\mu,p,t)$.

In modeling time profiles of solar particle events with solutions of the
transport equation with the goal to derive transport parameters, it is 
important to not only fit the isotropic part of the distribution function but
also make use of the information contained in its angular dependence. A first-order anisotropy parallel to the magnetic field can be defined as 
\begin{equation}
A_{p1} = \frac{3 \int_{-1}^{+1} d\mu\, \mu\, f(\mu)}
                {\int_{-1}^{+1} d\mu\, f(\mu)}.             
\end{equation}
If the scattering is sufficiently strong, $f(\Vec{r},\mu,t)$ adjusts rapidly to 
a nearly isotropic distribution, and the solutions of Equation~(2) become similar
to those of a spatial diffusion model. The mean free path $\lambda _{\parallel}$
which relates the pitch angle scattering rate to the spatial diffusion parallel 
to the ambient magnetic field is given by
\begin{equation}
\lambda _{\parallel} \ = \ {{3v}\over{8}} \int\limits_{-1}^{+1}
d\mu \ {{(1-\mu^2)^2}\over {D_{\mu\mu}(\mu)}}. 
\end{equation}
The parallel mean free path, related to the parallel diffusion coefficient by
$K_{\parallel} = v/3\cdot\lambda _{\parallel}$ is often used as a convenient parameter to characterize the varying degrees of scattering from one solar particle event to another, 
even when it adopts values close to or larger than the observer's distance 
from the Sun and the transport process can not be considered as spatial
diffusion.\\

\begin{figure} 
 \centerline{
 \includegraphics[width=0.5\textwidth,clip=true]{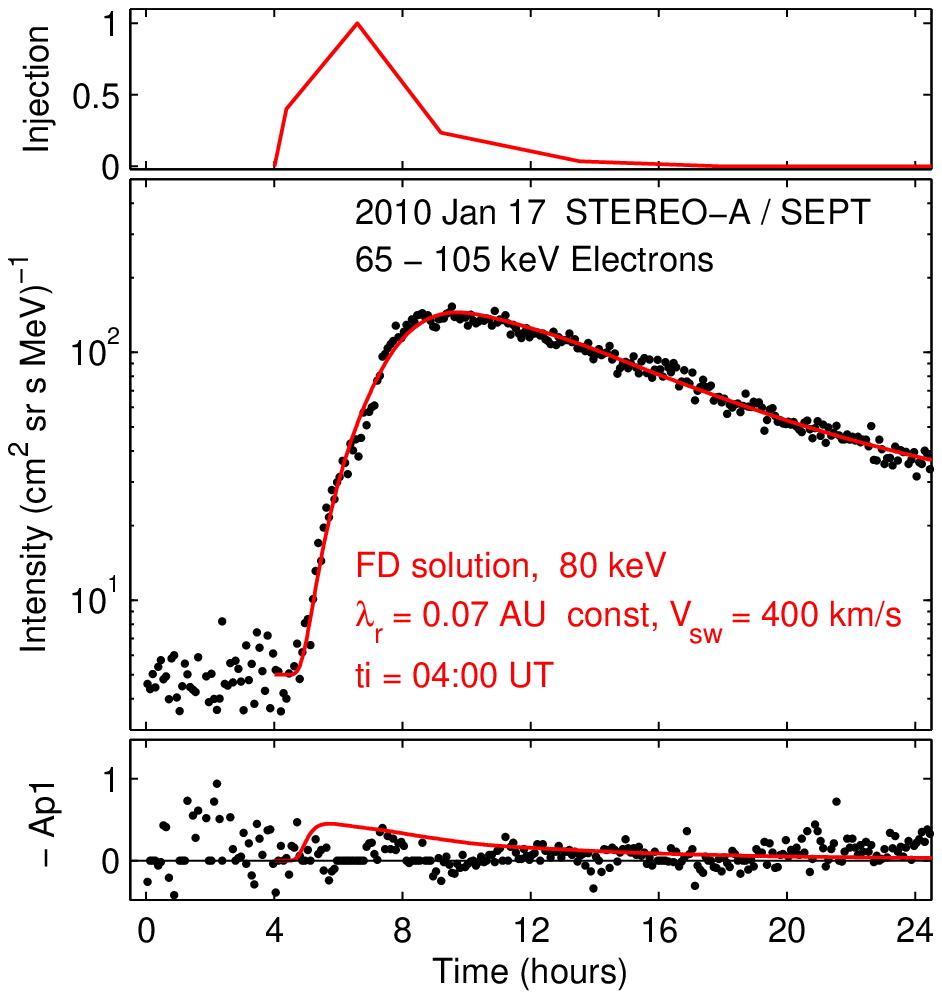}
 \includegraphics[width=0.5\textwidth,clip=true]{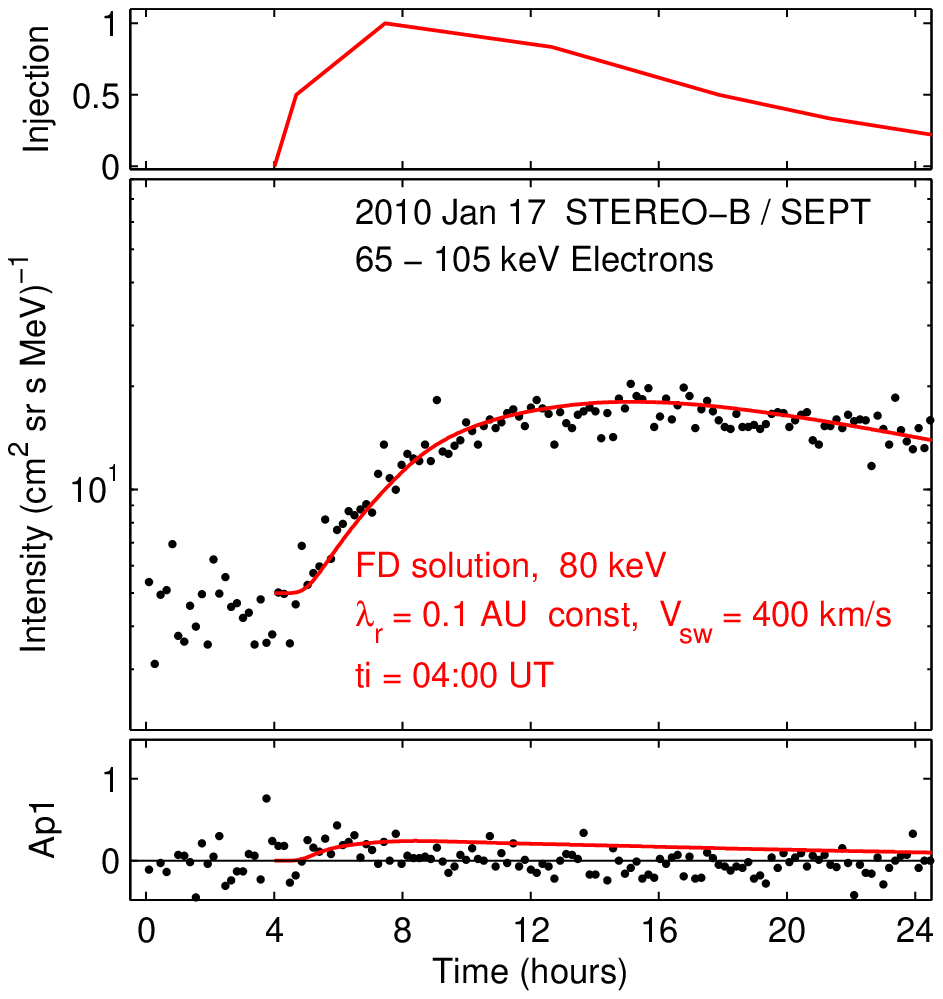}}
 \caption{Fits obtained from the FD model to the omnidirectional intensity-time and anisotropy-time profiles of 65-105\,keV electrons observed by STEREO~A (left) and STEREO~B (right). The top panel show the derived injection functions.}\label{fig:sim1}
 \end{figure}

Figure \ref{fig:sim1} (left) shows a fit obtained from the FD model to the 
65-105\,keV electron observations on STEREO~A. Unfortunately, at the
onset of the event, when the anisotropy usually reaches a maximum, the
pitch angle coverage of the instrument was not favorable (see Fig. \ref{fig:ani}). No clear 
anisotropy profile which could be used for the modeling could be derived.
The data indicate that the anisotropy was small throughout the event, not
exceeding a value of 0.4 or so. Under this restriction, a reasonably good
fit was achieved assuming a radially constant mean free path of  
$\lambda_r$ = 0.07 AU and the injection profile shown in the upper panel.

%
%

Similarly, a fit to the electron observations on STEREO~B was obtained 
assuming a radially constant mean free path of $\lambda_r$ = 0.1 AU.
The injection profiles derived for the STEREO~A and STEREO~B observations
are much longer than what is usually observed for electrons in this
energy range. 
But because of the large angular distance the particles have to propagate in the 17 January 2010 event, and which is, according to this model, in the solar corona, these findings might not be unrealistic.

\subsection{Model 2 - Perpendicular Diffusion in the IP Medium}
For an alternative interpretation, we consider a second model which assumes
that the particles are released from a source region with a width of 20
degrees in longitude and latitude, centered at the location of the optical
flare. No coronal propagation is assumed, instead any lateral transport occurs
due to diffusion of the electrons in the interplanetary medium perpendicular
to the ambient magnetic field. The model is based on a Monte-Carlo solution 
of the stochastic differential equations corresponding to Equation~(2) (cf., \inlinecite{Droege2010}). We still neglect energy 
losses and convection of the electrons with the solar wind, but now take 
into account perpendicular diffusion and co-rotation. Accordingly, the SDE 
for the momentum transport is omitted and only the following two SDE are left:
\begin{eqnarray}
d\Vec{r}(t)=\mu \upsilon \Vec{b}\:dt + \sqrt{2\Vec{K_{\bot}}}~d\Vec{W_{\bot}}(t)
+ {\mbox{\boldmath $\nabla$}} \Vec{K_{\bot}}dt, \hspace*{0.4cm}
\end{eqnarray}\\[-5.2ex]
\begin{eqnarray}
d \mu(t) = \sqrt{2 D_{\mu\mu}} \ dW_{\mu}(t) + \biggl[{v \over 2 L} (1 - \mu^2)
+  {\partial D_{\mu\mu}\over\partial \mu} \biggr] dt, \
\end{eqnarray} 
where $W_{\mu}(t)$ and $\Vec{W_{\bot}}(t)$ denote one- and two-dimensional Wiener processes, respectively.
 \\
\begin{figure} 
 \centerline{
 \includegraphics[width=0.5\textwidth,clip=true]{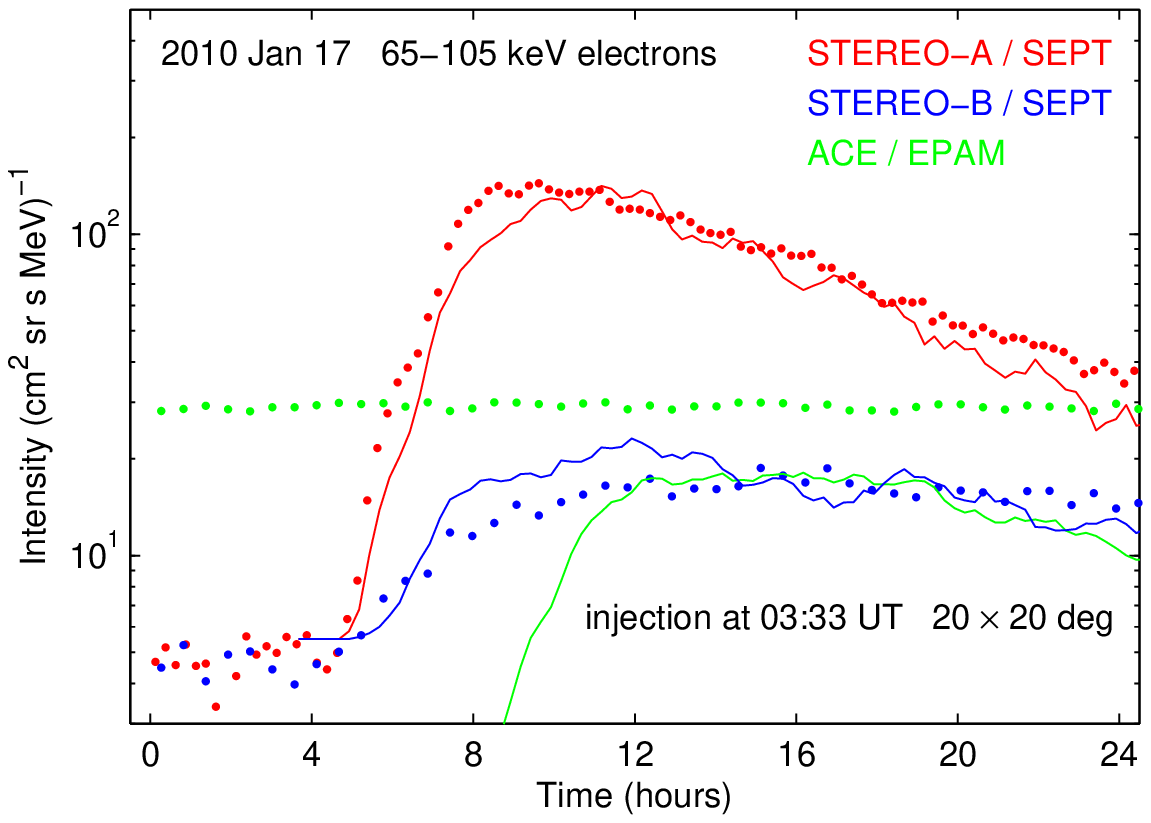}
 \includegraphics[width=0.5\textwidth,clip=true]{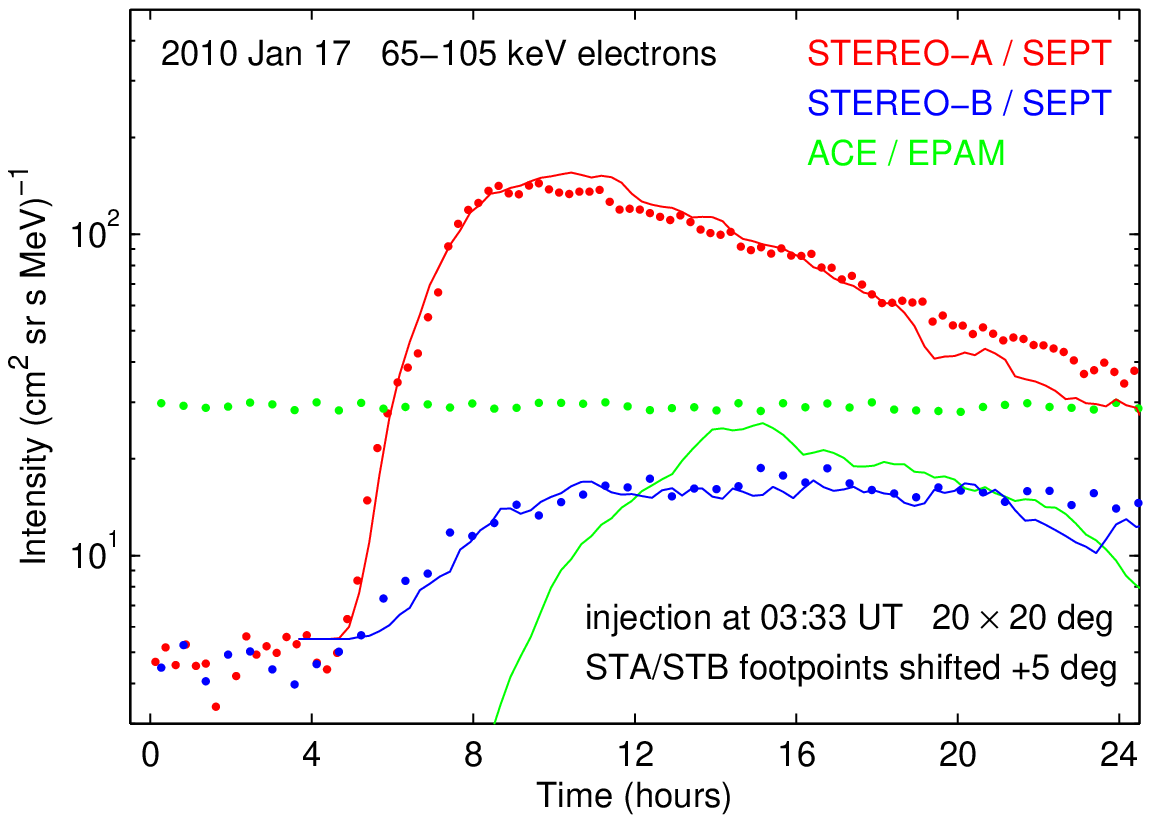}}
 \caption{Left: Fits obtained from the MC model to the omnidirectional intensity-time profile of 65-105 keV electrons observed by by STEREO~A and
STEREO~B assuming azimuthal distances of the footpoints from the flare 
region derived from the observed solar wind speeds. Right:  Fits obtained 
assuming that the footpoints are shifted in the western direction by
5 degrees.}\label{fig:sim2}
\end{figure}%

Following our previous work, we consider that the perpendicular mean free path scales with the gyroradius of the particle, i.e., with the magnetic field strength and with the particle's pitch angle \cite{Droege2010}:
\begin{equation}
\lambda_{\perp}(\mu,r) \ = \ \alpha \ \cdot \lambda_{\parallel}(r) \cdot  \left(\frac{r}{1\mbox{AU}}\right)^2 \cdot \cos(\psi(r)) \cdot \sqrt{1-\mu^2}, 
\end{equation}
where $\psi$ is the angle between the radial direction and the magnetic 
field, and $\alpha$ gives approximately the ratio $\lambda_{\perp}/\lambda_{\parallel}$ at 1 AU in the case that the electron distribution function is nearly isotropic at that location. We note that for smaller radial distances and anisotropic distribution functions 
$\lambda_{\perp}/\lambda_{\parallel}$ can be considerably smaller than
$\alpha$.

Figure \ref{fig:sim2} (left) shows the electron intensities observed on STEREO~A and B,
and in addition also ACE/EPAM electron observations in the same energy
range. It is evident from the figure that the ACE/EPAM background is considerably higher than on both STEREO spacecraft, and that a possible electron flux resulting from the flare stays below that background (c.f. Fig. \ref{fig:intro}).
However, the background flux was used as a constraint for the modeling,
the results of which are shown as solid lines in the figure. 
It was
assumed that electrons are injected at a radial distance 0.05 AU, with an 
energy distribution according to a power law with a spectral index 
$\gamma$ = 3. 
Note, that the model assumes an impulsive injection of the particles. The injection hight of 0.05\,AU has been chosen to save computation time. However, a lower injection hight of i.e. 0.02\,AU as suggested by the highest frequency of the type III bursts would not change the model results at the existing resolution of the observations.
In accordance with the first model, a constant 
$\lambda_r$ = 0.1 AU was assumed, and a value of $\alpha$ = 0.3.
The parallel anisotropy in this event was not quantifiable but probably very low and was not used for the modeling.
The figure shows that the resulting intensities,
which were normalized by a single factor, reproduce basic features of the
electron observations. Both the shapes of the intensity profiles observed
on STEREO~A and B and the ratio of the maximum intensities are matched
well. The intensity predicted for the near-Earth environment stays well
below the observed background for ACE/EPAM. 
Note, that the SOHO/EPHIN electron observations, which show the SEP event, have not been modeled. 
Because of the much higher energy range (250-700\,keV) this would require too many additional assumptions like on the spectral shape and an energy dependence of the mean free path.

One of the uncertainties in the modeling is the exact structure of the interplanetary magnetic field and of the resulting extrapolation of the
location of the origin of the field line connecting to the spacecraft.
The right side of Figure \ref{fig:sim2} shows the result of a modeling where the
origins of the field lines have been shifted counter-clockwise (to the west)
by 5 degrees, and the other parameters have been left unchanged. As can be
seen, this moderate shift which is well within the limits of the
uncertainties regarding the distance between the source region and the connecting field lines leads to a nearly perfect fit of the intensity
profiles.

\section{Discussion of Model Results and Conclusions}
Although the source of flare related energetic particles is commonly treated as a point source, we applied an extended source in model 2. 
This is suggested by the remote observation of a shock through the type II radio burst (c.f. \inlinecite{Grechnev2011}).
On the other hand, Potential Field Source Surface (PFSS) models, describing the coronal magnetic field, also support an extension of the source region, which is due to diverging magnetic field lines as described by \inlinecite{Klein2008}. 
\begin{figure} 
 	\centerline{
 	\includegraphics[width=0.33\textwidth]{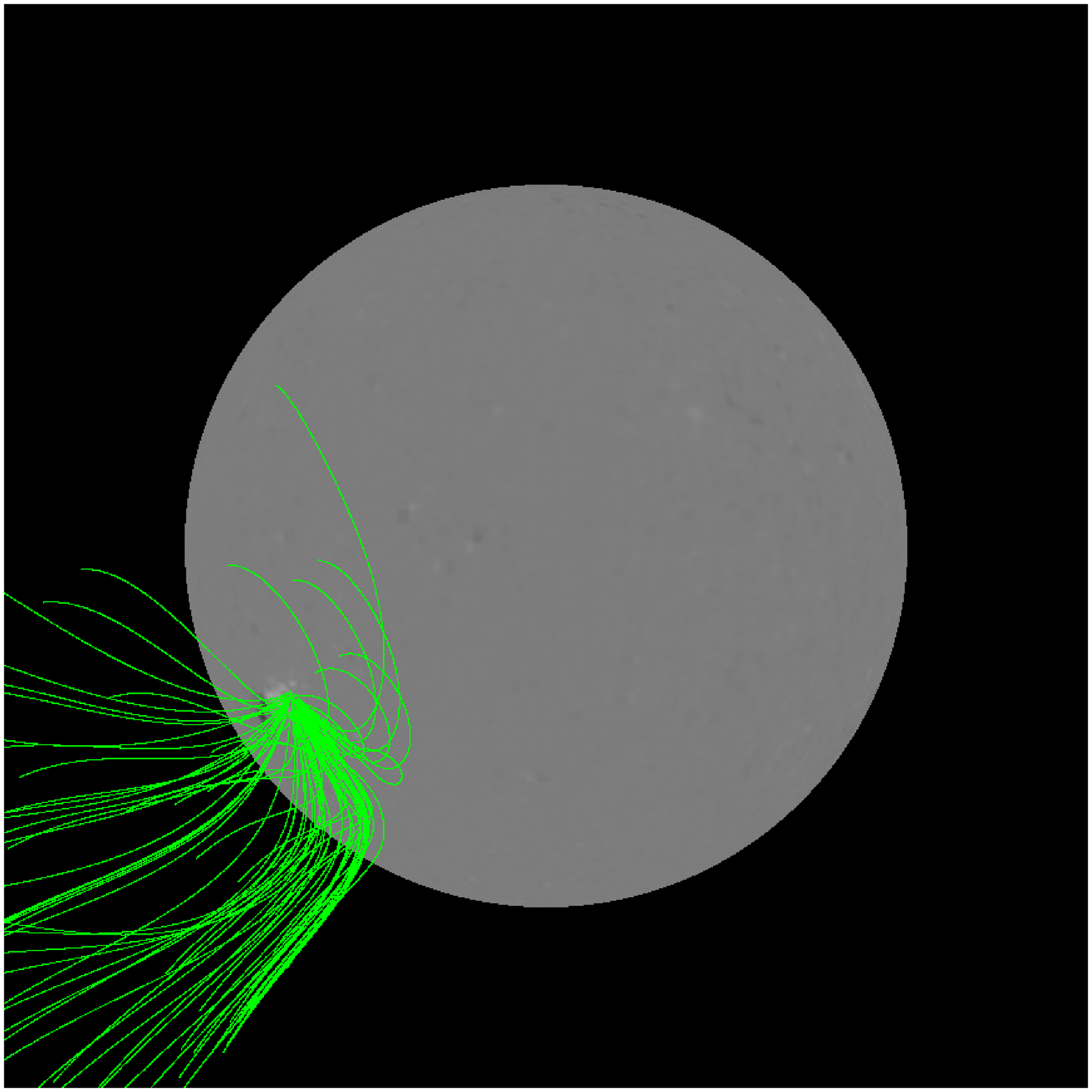}
 	\includegraphics[width=0.33\textwidth]{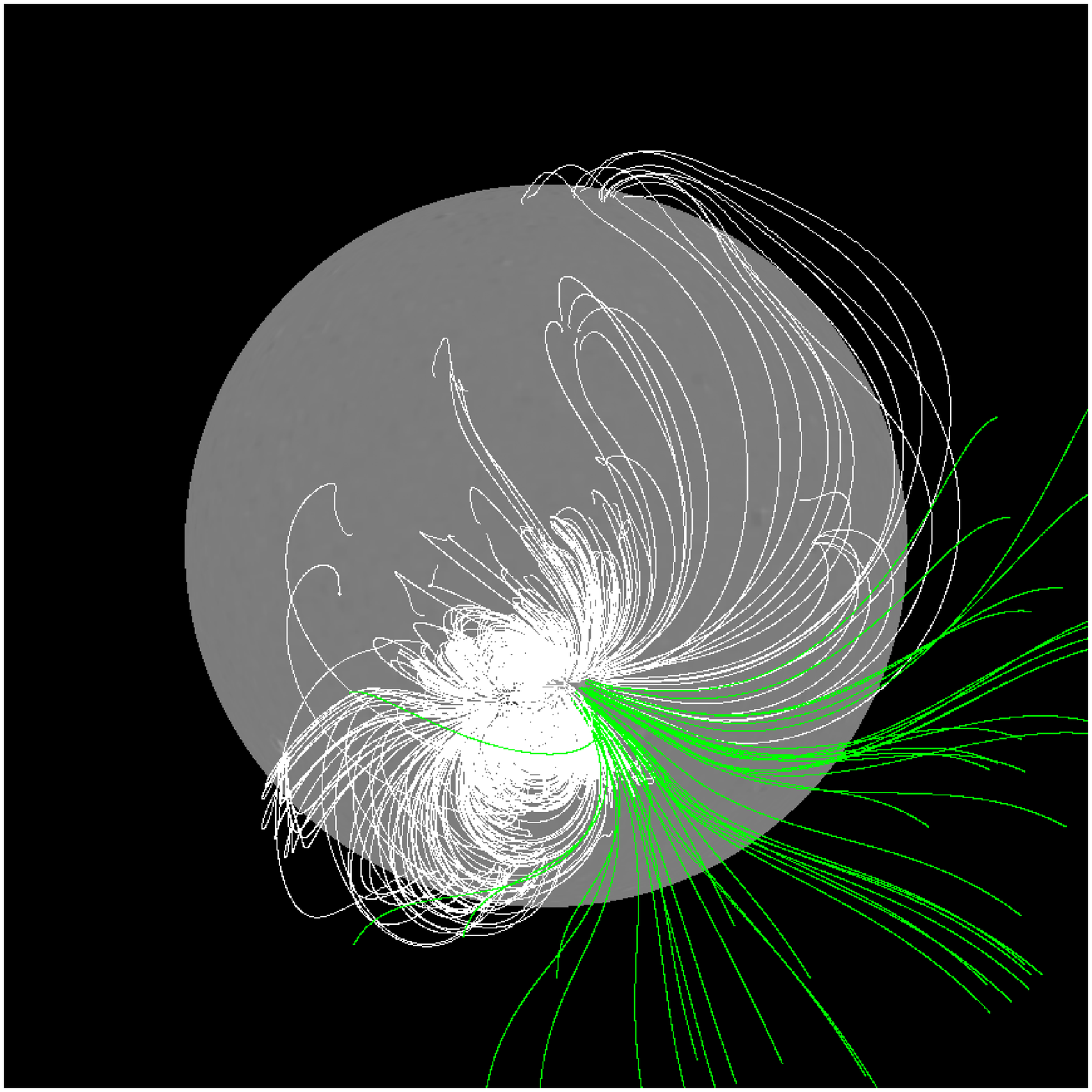}
 	\includegraphics[width=0.33\textwidth]{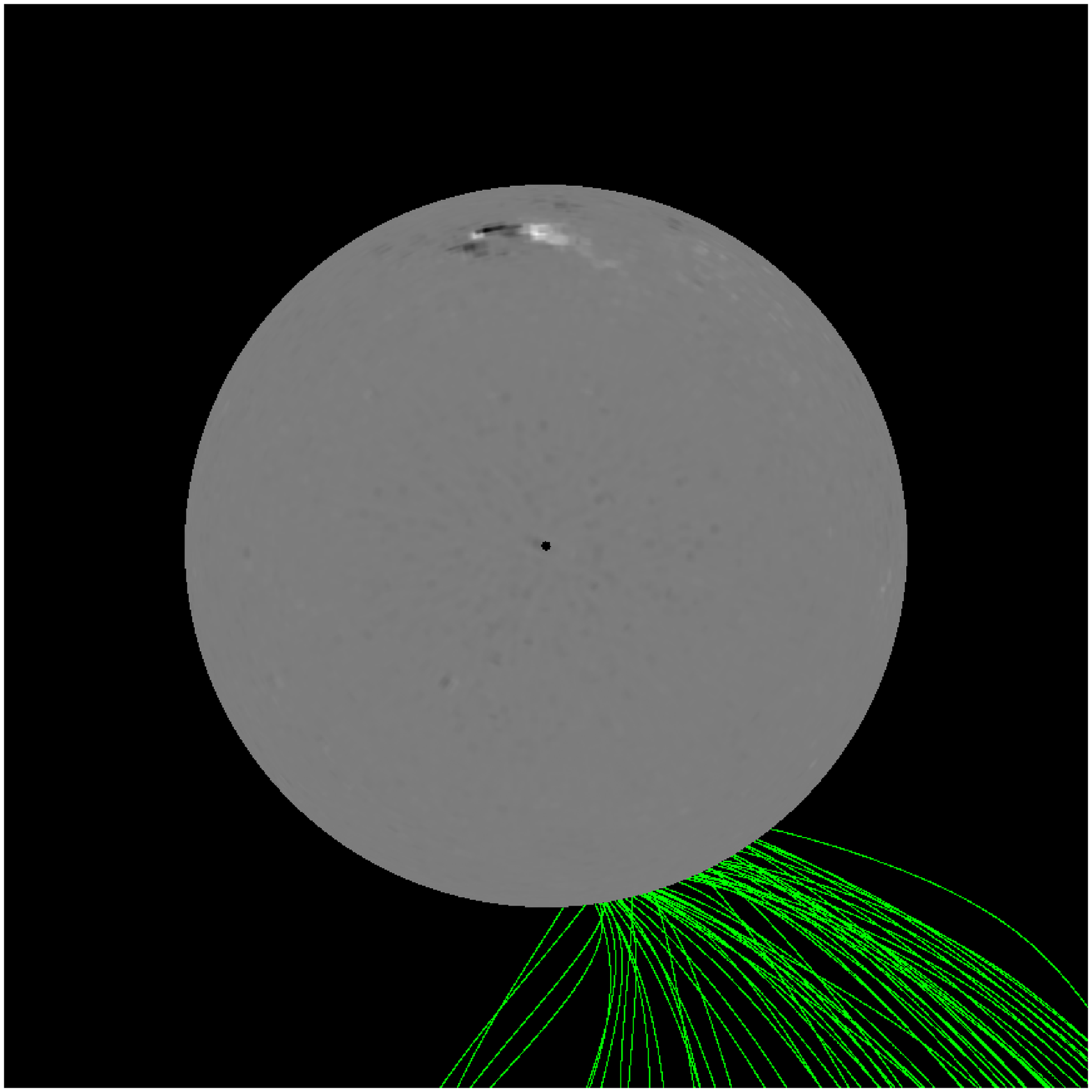}
 	}
	\caption{Potential Field Source Surface (PFSS) plots produced with the solarsoft package pfss\_viewer. The figures show open field lines (green) of positive
	 polarity and closed field lines (white, only center figure) emerging from the AR, which produced the flare on 17 January. The left figure shows the 17 January STEREO~B perspective, the center figures presents a frontal view onto the AR, and the left figure shows a view from the north.}\label{fig:pfss}
\end{figure}%
Figure \ref{fig:pfss} shows three different views to the flaring active region of the presented event. 
The green lines show open magnetic field lines of positive polarity emerging from the positive part of the active region. 
The white lines represent closed field lines and are only shown in the center panel of the figure. 
The central viewing angle for the left figure is 120 degrees, the center figure shows the AR more frontal with an angle of 70 degrees, and the right hand figure presents a view from the north. 
The starting height was set to 1.0159 solar radii and the upper height is 2.5 solar radii. 
The model shows that the open magnetic field lines are located to the west and a divergence of open magnetic field lines of several tens of degrees can be seen. 
This agrees with the second model run of model 2 (see Figure \ref{fig:pfss} (right)), where the source region was shifted by five degrees to the west.
Note that the AR is behind the limb for the Earth on 17 January 2010, so the magnetogram observations taken for this PFSS model are not up to date and could have changed. 
\\

We conclude that both models considered here for the lateral transport
can explain the electron observations on STEREO~A and B following the 
flare on 17 January 2010. 
However, because we do not find evidence for unusually long injection functions at the Sun, as proposed by model 1, we tend to believe more in a scenario with strong perpendicular diffusion in the interplanetary medium as described by model 2.
Of course a combination of both processes, coronal transport and perpendicular diffusion in the IP medium is likely a more realistic description of the problem.
Furthermore, the EUV wave directed towards the magnetic footpoint of STEREO~B (see Fig. \ref{fig:flare} right) may also  influence the coronal particle propagation as suggested by \inlinecite{Rouillard2012}.
More information about the nature of the lateral transport of solar energetic particles can be expected if more 
simultaneous two- or three-spacecraft observation of particle events
with a measurable anisotropy become available.
%

\section{Summary}\label{s:sum}  
We present multi-spacecraft observations of the 17 January 2010 solar energetic particle event, which has been detected by three viewpoints provided by the two STEREO spacecraft and the close to Earth spacecraft SOHO. 
Although it is a backside event as seen from Earth and all magnetic footpoints of the three spacecraft are separated by more than 100 degrees from the flaring active region, energetic electrons and protons are observed at all these positions.
The longitudinal spread of energetic particles in this event is consequently suggested to be nearly 360 degrees at 1\,AU.
Remote observations of the flare and of an EUV wave have been obtained by STEREO~B only, while a CME and type III radio bursts were also observed by STEREO~A, SOHO, and WIND.
A type II radio burst, indicating the presence of a coronal shock, was recorded by STEREO~B and the Earth-based radio observatory HiRAS, while no associated interplanetary shock has been detected in-situ.
Energetic particles arrive at the spacecraft strongly delayed with respect to the flare onset.
Because the pitch-angle coverage for STEREO~A, which measures the most significant event, is rather poor, it is difficult to obtain the anisotropy of the event. Later phases during the event with better pitch-angle coverage, however, show only very small anisotropy.
The long time delay, the large angular spread and small anisotropy point to the fact that the particles are not directly streaming away from an extended region but undergo strong scattering in the interplanetary medium.
In order to characterize the event in terms of particle transport in the interplanetary medium including perpendicular diffusion, two model approaches have been considered. While the first model describes the parallel particle propagation along the magnetic field lines and neglects perpendicular motion to the average magnetic field \cite{Droege2003}, the second model is a 3D propagation code including perpendicular diffusion in the IP medium \cite{Droege2010}, where an extended source of 20 degrees in latitude and longitude has been applied.
We find both models capable of describing the observed time-intensity profiles observed at the STEREO spacecraft.
The first model assumes all lateral transport to be performed in the corona and requires injection functions at the Sun with durations of several hours, while the second model yields a large ratio of perpendicular to parallel diffusion of 0.3.
Because we do not find evidence for extremely prolonged particle injections at the Sun we favor the second model results for explaining the observations, although we can not exclude a combination of both processes, perpendicular diffusion in the IP medium and coronal transport.
Nevertheless, the results suggest that the ratio of perpendicular to parallel diffusion may vary among different solar events and large values or the ratio have to be considered.
%
%
\begin{acks}
We acknowledge STEREO PLASTIC, IMPACT, SECCHI, EIT and Wind teams for providing the data used in this paper. 
The STEREO/SEPT and SOHO/EPHIN projects are supported under Grant 50 OC 0902 by the German Bundesministerium f\"ur Wirtschaft through the Deutsches Zentrum f\"ur Luft- und Raumfahrt (DLR). 
We thank the Geophysical Institute, University of Alaska, Fairbanks for providing the ecliptic plane IMF maps.
The radio monitoring survey is generated and maintained at the Observatoire de Paris by the LESIA UMR CNRS 8109 in cooperation with the Artemis team, Universities of Athens and Ioanina and the Naval Research Laboratory. 
This work utilizes data obtained by the Global Oscillation Network Group (GONG) program, managed by the National Solar Observatory, which is operated by AURA, Inc. under a cooperative agreement with the National Science Foundation. 
The data were acquired by instruments operated by the Big Bear Solar Observatory, High Altitude Observatory, Learmonth Solar Observatory, Udaipur Solar Observatory, Instituto de Astrof\'isica de Canarias, and Cerro Tololo Interamerican Observatory.  
Y.K. was partially supported by the Russian Foundation for Basic Research (project No. 09-02-00019-a).\\
\end{acks}
%
%
%
\bibliographystyle{spr-mp-sola}
\bibliography{jan17_ref}  
\end{article} 
\end{document}